\documentclass[final,3p,times,twocolumn]{elsarticle}
\usepackage{graphicx}
\usepackage{amssymb}
\usepackage{gensymb}
\usepackage{natbib}
\usepackage{mathtools}
\usepackage{bm} 
\usepackage{booktabs}

\usepackage{lineno}

\begin{document}

\begin{frontmatter}

\author[label1]{S. B. Harris}
\author[label1]{J. H. Paiste}
\author[label2]{J. Edoki }
\author[label3]{R. R. Arslanbekov}
\author[label1]{R. P. Camata}
\ead{camata@uab.edu}

\address[label1]{Department of Physics, University of Alabama at Birmingham, Birmingham, Alabama 35294, USA}
\address[label2]{Alabama A\&M University, Normal, Alabama, 35762, USA}
\address[label3]{CFD Research Corporation, Huntsville, Alabama, 35806, USA}

\begin{abstract}
We carry out simulations of laser plasmas generated during UV nanosecond pulsed laser ablation of the chalcogens selenium (Se) and tellurium (Te), and compare the results to experiments. We take advantage of a 2D-axisymmetric, adaptive Cartesian Mesh (ACM) framework that enables plume simulations out to centimeter distances over tens of microseconds. Our model and computational technique enable comparison to laser-plasma applications where the long-term behavior of the plume is of primary interest, such as pulsed laser synthesis and modification of materials. An effective plasma absorption term is introduced in the model, allowing the simulation to be constrained by experimental time-of-flight kinetic energy distributions. We show that the effective simulation qualitatively captures the key characteristics of the observed laser plasma, including the effect of laser spot size. Predictions of full-scale experimentally-constrained Se and Te plasmas for 4.0 J/cm$^2$ laser fluence and 1.8 mm$^2$ circular laser spot area show distinct behavior compared to more commonly studied copper (Cu) plumes. The chalcogen plumes have spatial gradients of plasma density that are steeper than those for Cu by up to three orders of magnitude. Their spatial ion distributions have central bulges, in contrast to the edge-only ionization of Cu. For the irradiation conditions explored, the range of plasma temperatures for Se and Te is predicted to be higher than for Cu by more than 0.50 eV.
\end{abstract}

\title{Experimentally constrained multidimensional simulation of laser-generated plasmas and its application to UV nanosecond ablation of Se and Te}

\journal{arXiV}
\end{frontmatter}
\raggedbottom
\section{Introduction}
\label{S:1}
The complexity and broad technical applications of laser-generated plasmas have long been motivators for their study via numerical simulations \cite{Luchin1983,Vertes1989,Balazs1991}. Modern studies based on a variety of fluid \cite{Chen2005,Zhang2001,Aghaei2008,Autrique2013}, kinetic \cite{Han2002,Pietanza2010,Stokes2021}, and hybrid \cite{Itina2002,Palya2018} models have been well described in the literature. The majority of investigations have focused on one-dimensional (1D) descriptions of single-element plumes, very close to the target surface (\textless 1 mm) and over short timescales (\textless 1 $\mu$s). Detailed analysis of this domain is valuable in applications dominated by near-target processes, such as micromachining \cite{Tani2008}, laser-induced breakdown spectroscopy (LIBS) \cite{Nosrati2020}, and laser sintering/additive manufacturing \cite{Bolduc2018}. Simulations up to several centimeters in length and times \textgreater 10 $\mu$s, have received much less attention but are useful to guide materials synthesis and modification approaches, such as pulsed laser deposition (PLD), where substrates are placed several centimeters away from ablation targets \cite{Harris2019b}. 

Simulations over long distances are computationally costly and must include three-dimensional (3D) effects. While embedding detailed mechanisms is desirable, codes of practical utility need to be fast and meet predictive benchmarks. In this work we balance these demands using an effective simulation model. The model incorporates well established features of laser ablation plasmas, including thermal evaporation due to laser-solid interaction, bremmstrahlung emission from plasma electrons, as well as photoionization and free-free laser absorption by the plasma. An additional effective plasma absorption coefficient is introduced as a stand-in term for more sophisticated mechanisms that are impractical to account for directly\textemdash due to computational cost or unavailability of physical parameters. The effective model is constrained by Langmuir probe experimental data. We determine the value of an adjustable parameter\textemdash the pre-factor of the effective absorption coefficient\textemdash that causes simulated kinetic energy distributions to be consistent with their experimental counterparts. Once the optimum value of the adjustable parameter has been obtained, full scale simulations of elemental plumes are performed. The simulations are set up in a 2D-axisymmetric geometry employing a solver with an adaptive Cartesian mesh (ACM).

We use the effective model to simulate laser-generated plasmas of the chalcogens selenium (Se) and tellurium (Te), which are of current interest in laser synthesis of transition metal mono- and dichalcogenide materials. Research on these compounds has significant potential for uncovering new physics and enabling novel devices. Ultrathin monochalcogenide $\beta$-FeSe on SrTiO$_3$, for example, exhibits high-$T_c$ superconductivity \cite{Ge2015} and giant thermoelectric power factor above $T_c$ \cite{Shimizu2019}. Growth of this material via plasma-mediated approaches may allow interface-enhanced superconductivity in multilayer systems \cite{Coh2016} and electric-field controlled devices \cite{Shiogai2016}. Dichalcogenides like WSe$_2$ and MoTe$_2$ are, in their own right, part of a class of 2D materials with unique optoelectronic properties. They have implications for the future of flexible LEDs \cite{Sundaram2013}, photovoltaics \cite{Tsai2014}, 2D transistors \cite{Akinwande2019}, and quantum information processing devices \cite{Jha2018}. It has been demonstrated that an elemental chalcogen plasma can be used to alloy transition metal dichalcogenides \cite{Mahjouri-Samani2015}. Indeed, techniques based on laser-generated plasmas are emerging as a viable approach for the synthesis and processing of 2D materials \cite{Mahjouri-Samani2014}. Our effective model may be useful to tailor specific properties of laser-generated plasmas. This may allow a level of kinetic control in materials growth and processing that is inaccessible in chemical vapor deposition and vapor phase transport techniques \cite{Puretzky2020}.

\section{Model Description}
\label{S:2}
\subsection{Preliminaries}
The spatiotemporal evolution of laser-generated plasmas is a complex phenomenon that couples thermodynamic, electromagnetic, and quantum mechanical processes. Progress in tackling this problem has been made over the past two decades by the implementation of increasingly sophisticated models, enabled by advances in computational techniques and resources \cite{Chen2005,Zhang2001,Aghaei2008,Autrique2013,Han2002,Pietanza2010,Stokes2021,Itina2002,Palya2018,AitOumeziane2014,AitOumeziane2018}. In broad strokes, simulation of the plasma formed upon laser irradiation of a solid requires the integration of processes of ejection of ablated species from the material, with processes of plasma plume ignition and expansion. These events occur with significant mutual interference. Processes of material removal involve light-matter interactions that depend on the laser wavelength, pulse duration, and target material properties. Thermal evaporation, subsurface boiling, supercritical boiling, as well as electronic and hydrodynamic sputtering vary in their relative importance in multiple regimes \cite{KellyandMiotello}. It is generally assumed that for sufficiently low laser fluence, thermal evaporation models are adequate for describing ablation by nanosecond pulses \cite{Marla2011}, whereas additional mechanisms need to be considered in the high fluence regime \cite{Yoo2000,Bulgakova2001}. If temperatures approach or exceed the critical temperature of the target, for example, supercritical processes take place. This requires models that include both, surface and volume removal mechanisms \cite{Autrique2013}. In all cases, the physical properties of the target, such as reflectivity, absorption coefficient, and density, develop temperature and time dependencies, which need to be taken into consideration \cite{Lutey2013,Marla2014}. Once the material has been ejected, models of plasma formation and expansion are needed. Numerous collisional and radiative processes, including single- and multi-photon absorption, impact ionization, electron impact excitation/deexcitation, inverse bremsstrahlung, and plasma reflection may affect the electron and ion populations \cite{Autrique2013a}. Laser plasmas may also contain molecular clusters and nanoparticles, calling for inclusion of molecular absorption and scattering processes \cite{Rozman2008}. The  evolution of the plasma then needs to be calculated using kinetic or fluid models. For fluid models, local thermodynamic equilibrium \cite{Marla2011} or two-temperature assumptions are generally required for practical simulations \cite{Autrique2013a,Oumeziane2016}. Beyond fluid descriptions, electromagnetic effects are also known to be significant. The emission of fast electrons prior to lattice melting, for example, leaves the target with a positive charge \cite{Amoruso1999}. Space charge caused by this emission, as well as ion and prompt electron ejection from early plasma formation, has been implicated in the acceleration and broadening of the kinetic energy distribution of ions \cite{Farid2013,Haider2017}. Analysis of the plasma expansion may therefore need space charge corrections. Finally, geometrical effects can also modify the plasma evolution. Changes in laser spot size alter the interaction volume between the laser pulse and the initial vapor, and may also affect the efficiency of evaporation due to heat dissipation changes in the target \cite{Stokes2021,Ranjbar2020}. Competing effects lead to changes in the forward peaking of the expansion and can result in important variations of the plasma plume angular distribution \cite{Saenger1994,Toftmann2003}. 

This broad variety of processes presents a challenge to plasma modellers. The inclusion of more intricate mechanisms in 3D simulations have high computational cost and generally involves the introduction of parameters whose values are experimentally unknown and whose \emph{ab initio} computation is impractical or involve large uncertainties. While the gradual incorporation of more refined processes will continue to yield useful insights as computational power grows, in this paper we employ an effective model that allows fast predictive simulations to guide materials processing. This approach acknowledges the difficulty of accurately capturing all relevant processes involved in laser ablation of arbitrary materials. Instead, we adopt a model with basic well-known process constituents, and incorporate an additional effective mechanism in the laser-plasma interaction\textemdash
which is a convenient entry point. The high computational efficiency permits a broad sweep on the effective parameter introduced in the model. Its appropriate value for specific irradiation conditions and target material is determined by constraining basic simulation outcomes to Langmuir probe experimental data. This approach is compatible with introduction of other effective parameters related to mass removal process from the target or further aspects of laser-plasma interactions. This method for studying laser plasmas, heretofore inhibited by the need for numerous recurrent and time-consuming simulations, is enabled by our efficient ACM implementation framework.

\subsection{Model details}
The effective model couples laser-induced surface evaporation with fluid plasma expansion based on previous formulations \cite{Vertes1989,Balazs1991,Chen2005}. The pertinent equations are expressed in 2D-axisymmetric coordinates.

The target material is irradiated at normal incidence with a laser pulse of temporal profile 
\begin{equation}\label{supergauss}
    I_{laser}(t) = I_0 \exp\bigg[ -\bigg(
    \frac{(t-t_c)^2}{2 \sigma_t^2}
    \bigg)^P\bigg],
\end{equation}
with $t_c = 15.0$ ns, $\sigma_t = 9.05$ ns, and $P=7$, corresponding to a pulse width of $\sim$25 ns. The value of $I_0$ is used to set the peak irradiance desired for each simulation. The laser intensity on the target surface $I(\mathfrak{r},\mathfrak{z},t)$, is centered at the origin of the radial coordinate $\mathfrak{r}$, with a radial fall-off for growing $\mathfrak{r}$ prescribed by the function   
\begin{equation}\label{beer}
    I(\mathfrak{r},\mathfrak{z},t) = I_{laser}(t) \exp{\bigg[ -\bigg(\sqrt{\mathfrak{r}/R_0}\bigg)^A} \bigg] (1-\mathcal{R}) e^{-\alpha \mathfrak{z}}
\end{equation}
with $A=12$ and $R_0$ representing the laser spot radius. The interaction of the laser pulse with the subsurface region of the target is modeled by the Beer-Lambert factor included in Eq. \ref{beer}, where the material's absorption coefficient is denoted as $\alpha$. A fixed target reflectivity $\mathcal{R}$ is assumed.

Heat transfer from the absorption volume into the bulk of the target is computed by solving the heat diffusion equation 
\begin{equation}\label{heat}
    \frac{\partial T(\mathfrak{r},\mathfrak{z},t)}{\partial t} = \nabla \cdot \bigg[\frac{\kappa}{(C_{p}\rho_m)} \nabla T(\mathfrak{r},\mathfrak{z},t)\bigg] + \frac{\alpha}{C_{p}\rho_m} I(\mathfrak{r},\mathfrak{z},t)
\end{equation}
where $\kappa$ is the thermal conductivity, $C_p$ is the specific heat, and $\rho_m$ is the mass density of the target material.

The pressure of the vapor that forms at the surface of the target $p_{vap}$ is calculated from the Clausius-Clayperon equation
\begin{equation}\label{claus}
    p_{vap}(T_{s}) = p_{0} \exp\bigg[\frac{\Delta H_{lv} (T_{s} - T_{b})}{RT_{s}T_{b}} \bigg],
\end{equation}
where $\Delta H_{lv}, T_b$, and $T_s$ stand for the heat of vaporization, boiling point, and surface temperature, respectively, and $R$ is the ideal gas constant. The vapor density $\rho_{vap,s}$ at the surface is calculated using the ideal gas law
\begin{equation}\label{idealgas}
    \rho_{vap,s} = \frac{p_{vap}}{kT_{s}}.
\end{equation}
Assuming a Maxwellian velocity distribution, the escaping atoms have normal velocity components whose mean is given by 
\begin{equation}\label{velocity}
    \nu_{vap,s} = \sqrt{\frac{2kT_{s}}{\pi m}}.
\end{equation}
Ionization of the vapor is governed by the set of Saha equations
\begin{equation}\label{Saha_1}
    \frac{X_e X^z }{X^{z-1}} = \frac{1}{n_{vap}} \bigg( \frac{2 \pi m_e k T}{h^2}\bigg)^{3/2} \exp\bigg( -\frac{IP^z}{kT} \bigg), \;z = 1,2
\end{equation}
where $X_e$ and $X^z$ are the fraction of electrons and ionic species with charge $z$, respectively. The number density of the vapor $n_{vap}$ is calculated from the mass density of the vapor $\rho_{vap}$ and the atomic mass. Ionization potentials for atoms with charge $z$ are represented by $IP^z$ and $T$ is the plasma temperature. Matter and charge conservation are enforced by $\sum_{z=0}^{2} X^z =1$ and $X^1+2X^2 = X_e$.

The laser irradiance that reaches the target surface is attenuated via a Beer-Lambert relation by absorption in the plasma. Four terms are included as contributions to this absorption: (i) electron-neutral and (ii) electron-ion inverse bremsstrahlung, (iii) single-photon photoionization, and (iv) an effective absorption process proportional to the number density of neutral species in the plasma. The latter is meant as a stand-in term, accounting for more complex processes of light absorption in the plasma, such as multi-photon ionization, and secondary effects such as electron impact excitation, whose cross-sections are generally unknown or difficult to compute \cite{Autrique2013a}. Because plasma absorption strongly impacts material removal from the target, this term may also provide an effective accounting of thermal processes beyond evaporation and non-thermal effects of laser-target interaction.

The total plasma absorption coefficient $\alpha_{absorb}$ may therefore be written as
\begin{equation}\label{alpha_absorb}
 \alpha_{absorb} = \alpha_{IB,e-n} + \alpha_{IB,e-i} + \alpha_{PI} + \alpha_{eff},
\end{equation}
where $\alpha_{IB,e-n}$ and $\alpha_{IB,e-i}$ are the absorption coefficients for electron-neutral and electron-ion inverse bremsstrahlung, respectively, $\alpha_{PI}$ corresponds to absorption by photoionization, and $\alpha_{eff}$ is the effective absorption coefficient.

The inverse bremsstrahlung terms are evaluated by \cite{Chen2005,Ranjbar2020}
\begin{equation}\label{e-nIB}
    \alpha_{IB, e - n} = \bigg[ 1- \exp\bigg( -\frac{h c}{\lambda k T}\bigg)\bigg] Q n_e n^0
\end{equation}
\begin{equation}\label{e-iIB}
\begin{multlined}
    \alpha_{IB, e - i} = \bigg[ 1- \exp\bigg( -\frac{h c}{\lambda k T}\bigg)\bigg] \frac{4 e^6 \lambda^3 n_e}{3 h c^4 m_e}\\
    \times \bigg( \frac{2 \pi}{3 m_e k T}\bigg)^{1/2} \sum_{z=1}^2 z^2 n^z 
\end{multlined}
\end{equation}
where $h$ is Planck's constant, $c$ is the speed of light in vacuum, $k$ is the Boltzmann constant, and $e$ is the elementary charge. The laser wavelength is represented by $\lambda$. The density of electrons and species with charge $z$ are $n_e$ and $n^z$, respectively. Ionization states up to $z=2$ are accounted for, which is expected  to be sufficient for relatively low laser irradiances. $Q$ is the cross section for photon absorption by an electron during a collision with a neutral atom. It has dependencies on the electron and photon energies, as well as on the momentum transfer cross section of the atom \cite{Autrique2014}. Within the scope of the effective model, we adopt a fixed value of $Q$ = 10$^{-40}$ cm$^5$ for all simulations. This represents an estimate of the mean value of $Q$ for the variety of temperatures and chemical species of interest in our study, at the laser wavelength used \cite{Autrique2014}.

The absorption coefficient for photoionization $\alpha_{PI}$ is computed by \cite{Ranjbar2020}
\begin{equation}\label{alphaPI}
    \alpha_{PI} = \sum_{z=0}^{2} \sum_{i=N_{min}^{z}}^{N_{max}^{z}} 
    x_i^z n^z \sigma_i^z(T)
\end{equation}
where $N_{min}^{z}$ is the minimum energy level of an atom with charge $z$ that can be ionized to charge $z+1$ by absorbing a single photon of energy 5 eV (248 nm), $N_{max}^{z}$ is the maximum energy level considered. The fractional occupation of level $i$ is calculated by $x_i^z = (g_i^z/U^z)\exp[-E^z_i/(kT)]$ where $U^z$ is the partition function of charge state $z$ and $g_i^z$ is the statistical weight of level $i$ and charge $z$. The photoionization cross section of level $i$ with charge $z$, $\sigma_i^z$, is calculated by \cite{Ranjbar2020}
\begin{equation}\label{sigma}
    \sigma_i^z(T) = \frac{32 \pi^2 (z+1)^2e^6 k_e^3}{3\sqrt{3} h^4 c \nu_{laser}^3 g_i^z}
    U^{z+1}(T) \frac{dE_i^z}{di}
\end{equation}
where $k_e$ is the Coulomb constant, $\nu_{laser}$ is the laser frequency, and $dE^z_i/di$ is the spacing between energy levels.

Finally, the effective absorption coefficient $\alpha_{eff}$ is assumed to be proportional to the number density of neutral species $n^0$, and expressed as
\begin{equation}\label{alpha_effective}
\alpha_{eff} =
  \begin{cases}
    0                & \quad \text{if } T < T_{min}\\
    C_{eff} n^0      & \quad \text{if } T > T_{min}\\
  \end{cases}
\end{equation}
where $T_{min}$ is a cut-off temperature that excludes regions too cold for the vapor to be characterized as a plasma. In the simulations presented in this paper $T_{min}$ was set to 1000 K.

The proportionality constant $C_{eff}$ is used as an adjustable parameter to constrain the model by experiments. The value of $C_{eff}$ for given irradiation conditions and target material is determined by matching the main features of simulated ion time-of-flight (TOF) kinetic energy distributions to experiments, as detailed in section \ref{Sub:2}. 

Expansion of the ablated material is described by the equations of hydrodynamics, representing conservation of mass (Eq. \ref{cons_mass}), momentum (Eq. \ref{cons_momentum}), and energy (Eq. \ref{cons_energy}), where $p$ is pressure, $\bm{\nu}$ is flow velocity, and $\rho$, $\rho \bm{\nu}$, and $\rho E$ are mass, momentum, and internal energy density, respectively. 

\begin{equation}\label{cons_mass}
    \frac{\partial \rho}{ \partial t} = - \nabla \cdot \big( \rho \bm{\nu} \big)
\end{equation}
\begin{equation}\label{cons_momentum}
    \frac{\partial \rho \bm{\nu}}{ \partial t} = -\nabla \big(p + \rho \bm{\nu}^2 \big)
\end{equation}
\begin{equation}\label{cons_energy}
\begin{multlined}
    \frac{\partial}{\partial t}\bigg[ \rho E + \frac{1}{2} \rho \bm{\nu}^2 \bigg] = - \nabla \cdot \bigg[ p \bm{\nu} + \bm{\nu}\bigg( \rho E + \frac{1}{2}\rho \bm{\nu}^2 \bigg) \bigg] \\ + \alpha_{absorb} I_{laser}(t) - \epsilon_{rad}
\end{multlined}
\end{equation}
In Eq. \ref{cons_energy}, energy is added to the plasma by $\alpha_{absorb}I_{laser}(t)$, with $\alpha_{absorb}$ given by Eq. \ref{alpha_absorb}, and is lost by bremsstrahlung radiation $\epsilon_{rad}$. Assuming a Maxwellian velocity distribution for the electrons, $\epsilon_{rad}$ is evaluated using \cite{Spitzer2006}
\begin{equation}\label{radiation}
    \epsilon_{rad} = \bigg( \frac{2 \pi k T}{3 m_e}\bigg)^{1/2} \frac{32 \pi e^6}{3 h m_e c^3} n_e \sum_{z=1}^2 z^2 n^z
\end{equation}
The pressure $p$ and temperature $T$ of the expanding plasma are related to $\rho$ and $\rho E$ by \cite{Zeldovich2002}
\begin{equation}\label{idealgas2}
    p = (1 + X_{e}) \frac{\rho k T}{m}
\end{equation}
\begin{equation}\label{internalenergy}
    \rho E = \frac{\rho}{m}\bigg[ \frac{3}{2}(1 + X_{e}) k T + IP^1 X^{1} + (IP^1 + IP^2) X^{2} \bigg]
\end{equation}
where  $X^1$ and $X^2$ are the fractions of singly and doubly ionized atoms. $IP^1$ and $IP^2$ denote the first and second ionization potentials of the atoms.

\section{Methods}
\label{S:3}
\subsection{Computational}
The simulation is carried out in a multidimensional setting using a state-of-the-art, open-source ACM framework \cite{basilisk.fr}. The compressible solvers available in this framework (Riemann and all-Mach solvers \cite{basilisk.fr}) were adapted to include the equations of state of the plasma (Eqs. \ref{Saha_1}, \ref{idealgas2}, and \ref{internalenergy}). The use of the dynamic ACM approach is essential since the initial ($t <$ 30 ns) spatial resolution required near the target is tenths of microns, while the simulation domain is larger by multiple orders of magnitude. As the plasma plume expands ($t >$ 30-60 ns), the fine near-target resolution is no longer required for most of the simulation domain which allows the grid to be coarsened. However, proper resolution of the moving plasma front is still necessary throughout the entire plasma dynamics until the front reaches the anticipated substrate location. Only the very narrow plasma front region needs to be resolved for this purpose, while the remainder of the computational mesh can stay coarse. The numerical efficiency of the ACM framework is further enhanced by an adaptive computation domain capability. This capability works in combination with ACM and allows for the size of the computational domain to automatically increase when the plasma front approaches its boundary. The technique is versatile (e.g., the domain size can be increased or decreased on demand by any predefined amount). In the present simulations, an expansion front nearing the computational domain boundary triggered the doubling of the domain size. Jointly with ACM, this method allows efficient and physically accurate simulations, starting from domain sizes of the order of millimeters and expanding to tens of centimeters, as needed. This combination of techniques thus makes the computation fast over the large distances from the target ($>$ 1-10 cm) and over long times ($t >$ 1-10 $\mu$s). The vacuum background is characterized by a fixed number density of neutral species, set to match the experimental conditions described in the next section.

\subsection{Experimental}
Measurements of laser plasmas analogous to the simulation were carried out inside a vacuum chamber with a background pressure of 3$\times$10$^{-6}$ torr. The focused beam of a KrF excimer laser (Lambda Physik LPX 305i), which has pulse duration of $\sim$25 ns and wavelength of 248 nm, was used to ablate solid metal targets at a 45$\degree$ angle of incidence. The laser spot size on the target was adjustable using rectangular apertures that blocked the periphery of the excimer laser beam. Apertures were placed in the beam path between the laser output window and the focusing lens. For all apertures used, the spot area on the target was approximately rectangular with a 3:1 aspect ratio. A 10 $\times$ 10 mm square, planar ion probe consisting of a thin molybdenum foil backed by an alumina plank was placed at a distance $d$ = 4.0 cm from the target surface, along the central axis of the plume. The probe was biased with $-70$ V, and the current resulting from the passage of the plasma pulse was determined from the voltage drop across a 10 $\Omega$ resistor, connected to ground.

Assuming quasi-neutrality conditions, the saturation current pulse $I(t)$ measured by the probe under strong negative bias, can be used to infer the electron density $n(t)$ at the probe location by \cite{Doggett2009}
\begin{equation}\label{current}
    n(t) = \frac{I(t)}{v A e},
\end{equation}
where $t$ is the time since the firing of the laser (i.e., the TOF of the fluid element being measured), $v$ is the plasma flow velocity, $A$ is the probe collection area, and $e$ is the electron charge. A practical approximation for $n(t)$ is obtained by using $v=d/t$ in Eq. \ref{current}, where $d$ is the distance between the ablation spot and the probe. 

Se and Te ablation targets were synthesized by pressing elemental powders (Alfa Aesar, 99.999\%) into 20 mm diameter discs. Each disc was sealed in an evacuated quartz ampoule (\textless 10$^{-3}$ torr) and sintered for 24 hours at 80\% of the melting temperature. The surface of each resulting high density, sintered target was polished before laser ablation. The Cu target was a nominally 99.999\% pure 1-inch diameter disc subjected to the same polishing step as the chalcogen targets.

\section{Results and Discussion}
\label{S:4}
Simulations for ablation of Se and Te were carried out with a laser fluence of 4.0 J/cm$^2$, which corresponds to setting  $I_0$ = 1.6$\times$10$^8$ W cm$^{-2}$ in Eq. \ref{supergauss}. For comparison purposes, results were also obtained for Cu, which is a commonly used representative metal in laser plasma studies.
\begin{table}[h]
\caption{Physical properties of Se, Te, and Cu adopted for the simulations.}
\centering
\resizebox{\linewidth}{!}{\begin{tabular}{@{}lllll@{}}
\toprule
\toprule
Physical Properties                                 & Se     & Te     & Cu\\ \midrule
Thermal conductivity $\kappa$ (W m$^{-1}$ K$^{-1}$)  & 0.519   & 1.97   & 401\\
Specific Heat $C_p$ (J kg$^{-1}$ K$^{-1}$)          & 321    & 202    & 385\\
Mass density $\rho$ (kg m$^{-3}$)                   & 4819   & 6240   & 8960\\
Boiling point $T_b$ (K)                             & 958    & 1261   & 2862\\
Absorption coefficient $\alpha$ ($\times10^7$ m$^{-1}$)          & 9.28 & 5.89 & 8.66\\
Reflectivity                                        & 0.40   & 0.22   & 0.37\\
First ionization potential $IP^1$ (eV)              & 9.75   & 9.00   & 7.73\\
Second ionization potential $IP^2$ (eV)             & 21.19  & 18.6   & 20.29\\
Heat of vaporization ($\times10^5$ J mol$^{-1}$)                 & 0.958 & 1.140 & 3.048\\ \bottomrule \bottomrule
\end{tabular}}
\label{tab:properties}
\end{table}

The adopted physical properties for the three chemical species are summarized in Table \ref{tab:properties}.
All listed properties were collected from the CRC Handbook of Chemistry and Physics \cite{CRC} except for the thermal conductivity values, which come from Lange's Handbook of Chemistry \cite{LNG}. The reflectivity and absorption coefficients are for 248 nm, and the values for Se and Te are the average of the single crystal values for parallel and perpendicular electric field orientations, to account for a polycrystalline target.
Atomic data required for calculating the absorption due to photoionization (not shown) was gathered from the NIST Atomic Spectra Database \cite{NIST_ASD} 

\subsection{General characteristics of plasma plumes}
\label{Sub:1}
\begin{figure}[t]
    \centering
    \includegraphics[width=\linewidth]{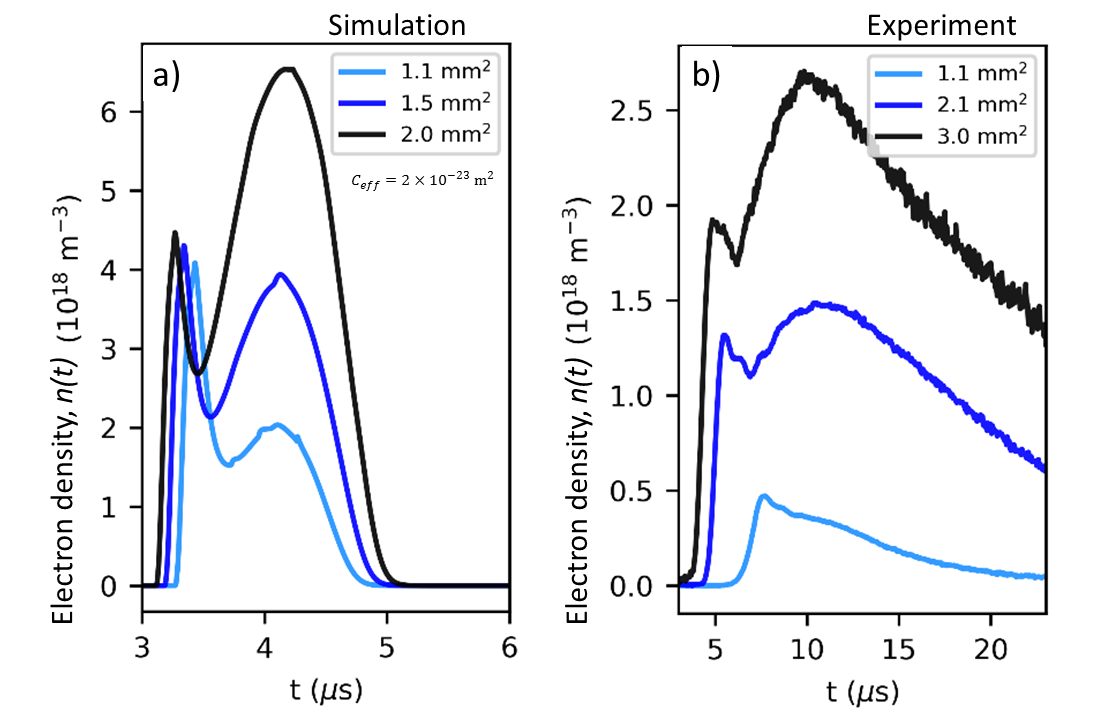}
    \caption{Effect of laser spot area in (a) simulated and (b) experimental electron density for tellurium (Te) ablation, at $d$ = 4.0 cm from the ablation spot, along the central axis of the plume. Simulation and experiment reveal similar trends in the relative amplitudes of fast and slow peaks with spot size variation. Calculations for all spot areas in (a) are for the arbitrarily chosen value of $C_{eff} = 2\times10^{-23}$ m$^2$, implying simulations unconstrained by experiments. Quantitative comparisons are therefore unwarranted. In both, simulation and experiment, the laser fluence was set to 4.0 J/cm$^2$.}
    \label{compare_spotsize}
\end{figure}

The electron density at a fixed position in space provides general characteristics of the plasma expansion. Its time dependence, $n(t)$, can be easily extracted from simulations and experiments. Fig. \ref{compare_spotsize}a shows simulated $n(t)$ for ablation of Te. The $n(t)$ traces are for a distance $d$ = 4.0 cm from the ablation spot, along the central axis of the plume. Simulations were run using an arbitrary $C_{eff}$ value. The curves show two peaks, with relative amplitudes that vary with the laser spot size. For the smallest spot size, a fast, high-density peak leads the expansion. A slower and broader peak of reduced density lags behind. As the spot size increases, the slow peak becomes dominant. As seen in Fig. \ref{compare_spotsize}b, the same trend is detected by the Langmuir probe in the experimental plasma\textemdash albeit over a broader range of spot size variation. Traces with comparable characteristics are also seen for ablation of Se (not shown). The similarity in trends between simulation and experiment indicates that the effective model can reproduce essential features of observed laser plasmas. Quantitative correspondence between simulation and experiment is not expected here, because of the arbitrarily chosen $C_{eff}$ value\textemdash Fig. \ref{compare_spotsize}a represents simulations unconstrained by experiments. It is particularly meaningful, however, that the 2D-axisymmetric simulation allows evaluation of the effect of laser spot size, which has not been widely studied, either experimentally \cite{Toftmann2003} or computationally \cite{Stokes2021,Ranjbar2020}. Spot size effects cannot be assessed in 1D simulations.

\subsection{Experimentally constrained simulations}
\label{Sub:2}
A simulation can be constrained by determining the value of $C_{eff}$ that causes the predicted $n(t)$ to be consistent with its experimental counterpart. For constraining considerations it is instructive to convert the variable $t$ into an energy axis defined by $\varepsilon=m (d/t)^2/2$, where $m$ is the mass of the corresponding ion. The quantity $\varepsilon$ is often referred to as the ``ion TOF kinetic energy'' \cite{Harris2019b}. 

\begin{figure}[!t]
    \centering
    \includegraphics[width =\linewidth]{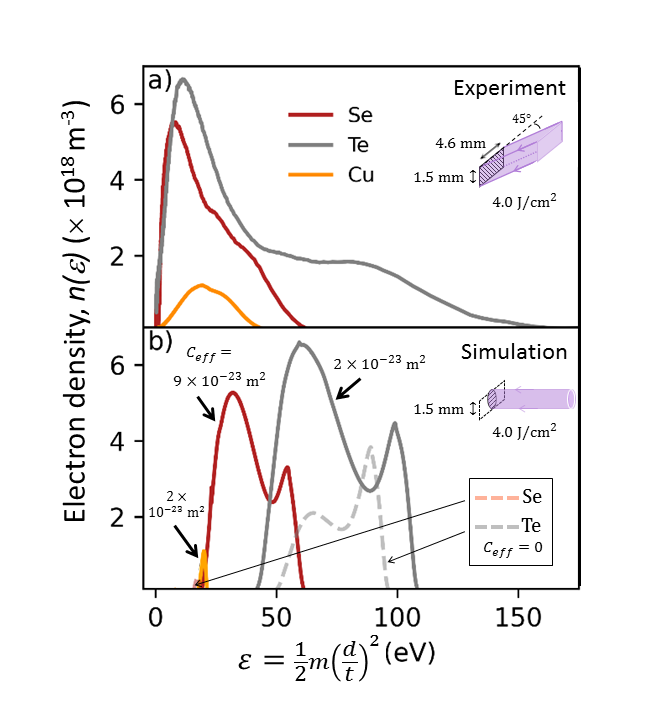}
    \caption{a) Experimental electron density $n(\varepsilon)$ for Se, Te, and Cu used to constrain the simulation. b) Simulated $n(\varepsilon)$ curves for approximately equivalent irradiation conditions: Dashed lines are predictions using $C_{eff}=0$; solid lines are for $C_{eff}$ values shown, which cause the maximum of each curve to match the maximum of the corresponding experimental $n(\varepsilon)$. Measurements and simulations are for $d$ = 4.0 cm from the ablation spot, along the central axis of the plume, and for a laser fluence of 4.0 J/cm$^2$. The insets illustrate the laser incidence on the target, with partial equivalence between experimental (rectangular, 1.5 mm $\times$ 4.6 mm) and simulation (circular, 1.5 mm diameter) laser spots.}
    \label{TOF}
\end{figure}

Fig. \ref{TOF}a shows experimental traces of $n(\varepsilon)$ for Se, Te, and Cu under the irradiation conditions illustrated in the inset. The Se plasma is dominated by ions with kinetic energy peaking at $\sim$7.6 eV, while a faster sub-population appears with energy in the 25-40 eV range\textemdash a tail of still higher energies extends up to $\sim$52 eV. The peaks for Te are broader and have greater peak energies compared to Se. The slow Te component peaks at $\sim$11 eV while the fast ions have maximum density at 79 eV. The peak ratios of the slow to the fast components are $\sim$1.8 for Se and 3.6 for Te. The $n(\varepsilon)$ curve for Cu is centered at $\sim$19.1 eV.

Simulated $n(\varepsilon)$ curves for approximately equivalent conditions are shown in Fig. \ref{TOF}b. Simulations using $C_{eff}$ = 0 lead to overall densities that are much lower than experiments, with values for Se and Cu off by more than one order of magnitude. In addition, the two-peak structure is noted to be inverted for Te. Fig. \ref{TOF}b also shows simulated $n(\varepsilon)$ using values of $C_{eff}$ that cause the maximum of each curve to match the maximum of the corresponding experimental trace. The best matches yield the values of $C_{eff}$ shown in the figure and in Table \ref{tab:C_eff}. While the width and exact peak locations retain substantial differences between simulation and experiment, we note the emergence of a two-peak structure that appears equivalent to the experimental chalcogen data. The peak ratios for both, Se and Te, evaluate to 1.6. This outcome matches the experimental peak ratio for Se, but is a factor of $\sim$2.2 lower than the Te measurement. The constrained Cu simulation produces a very narrow $n(\varepsilon)$ curve, that is centered at 19.3 eV. This is \textless 1\% 
different from the measured one. In the next section, we describe the simulation predictions when the $C_{eff}$ values determined by this constraining exercise are used throughout the calculations.

It is important to note that in applying the model to Se, Te, and Cu in this paper, we can only assert a partial equivalence between experimental and simulation environments. This is because of experimental limitations and computational symmetry requirements. The experimental laser spot is roughly rectangular due to the excimer laser electrode discharge geometry. The simulations, on the other hand, are axially symmetric, resulting in a circular laser spot on the target. The rectangular experimental spot implies a plume that deviates from axial symmetry. The amount of forward-peaking of the plume differs when viewed on the plane of the long side of the rectangle compared to the short side. By matching the simulation spot diameter to the short side of the experimental rectangle\textemdash as illustrated in the insets of Fig. \ref{TOF} \textemdash we are likely simulating a lower bound for the expansion velocities along the plume axis. Constraining the simulation by $n(\varepsilon)$ from more symmetric experimental conditions, may allow updated $C_{eff}$ values for improved predictions of overall plume behavior. It is also worth mentioning that the asymmetry of the experimental plume may be partially responsible for the broader experimental $n(\varepsilon)$ curves compared to the simulation.

\begin{table}[h]
\caption{Experimentally-constrained values of the pre-factor $C_{eff}$ (Eq. \ref{alpha_effective}) for a laser fluence of 4.0 J/cm$^2$ and circular simulation laser spot with 1.5 mm diameter.}
\begin{tabular*}{1.0\linewidth}{@{\extracolsep{\fill}}ccc}
\toprule
\toprule
Target Material & $C_{eff}$ (m$^2$)\\ \midrule
Se  & $9\times10^{-23}$\\
Te  & $2\times10^{-23}$\\
Cu  & $2\times10^{-23}$\\ \bottomrule \bottomrule
\end{tabular*}
\label{tab:C_eff}
\end{table}

\subsection{Simulation Results}
\label{Sub:3}
Full scale simulations employing the $C_{eff}$ values of Table \ref{tab:C_eff} can be conveniently described by dividing the temporal evolution of the plasma into three stages: laser-target interaction, early expansion and ionization, and long term expansion out to centimeter distances.

\begin{figure}[b!]
    \centering
    \includegraphics[width=\linewidth]{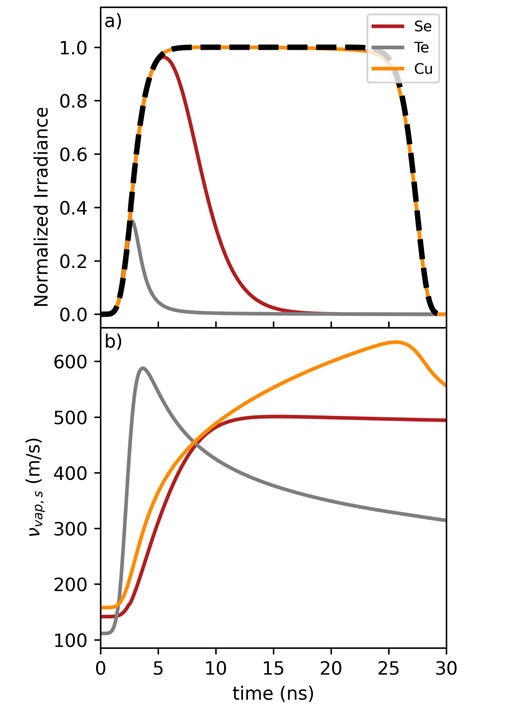}
    \caption{a) Normalized laser irradiance incident on the target, with black dashed line showing temporal profile of the laser pulse, prior to its interaction with the plasma. b) Mean normal velocity of vapor atoms during early plasma expansion. Simulations for laser irradiance of $I_0 = 1.6 \times 10^8$ W cm$^{-2}$, which corresponds to a 4.0 J/cm$^2$ fluence. Target irradiance values normalized by $I_0$.}
    \label{target comparison}
\end{figure}

The laser-target interaction and early expansion of the plumes (expansion distances much less than the spot size) exhibit very low sensitivity to the dimensionality of the description. 2D-axisymmetric and 1D results are essentially the same along the expansion direction. This allows us to explore these early processes at the reduced dimensionality of the 1D description with the finest spatial resolution, without the extra computational cost of multiple dimensions. 
Fig. \ref{target comparison}a shows the laser irradiance predicted to arrive at the targets for the chosen laser output fluence of 4.0 J/cm$^2$. In the case of Cu, very little plasma shielding occurs. Essentially all of the laser energy reaches the target throughout the duration of the pulse.

\begin{figure}[ht!]
    \centering
    \includegraphics[width = 0.9\linewidth]{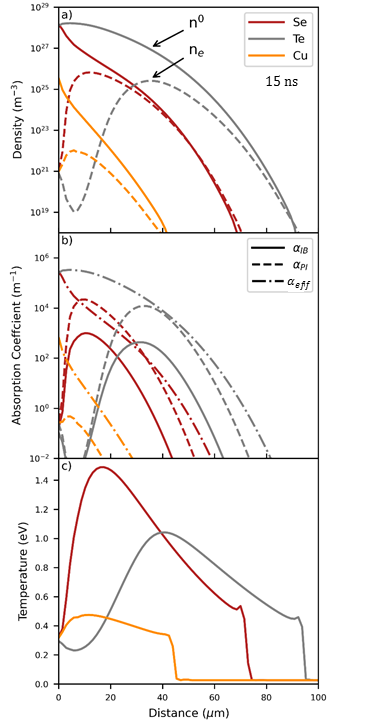}
    \caption{Detailed view of the density, plasma absorption coefficients, and temperature vs distance of the Se, Te, and Cu plumes at the midpoint time of the laser pulse, 15 ns. a) Density of neutral species $n^0$ and electrons $n_e$. b) Plasma absorption coefficients. Inverse bremsstrahlung absorption $\alpha_{IB}$ is the sum of the electron-neutral and electron-ion contributions, $\alpha_{PI}$ is absorption from single-photon photoionization, and $\alpha_{eff}$ is the effective absorption coefficient defined by Eq. \ref{alpha_effective}. c) Plasma temperatures peak near the position of greatest $\alpha_{PI}$.}
    \label{n0alphaT}
\end{figure}

For Se and Te, on the other hand, strong plasma shielding is predicted. For Se, the onset of plasma shielding occurs just as the laser reaches peak power density, while for Te, shielding begins at $\sim$35\% of the peak. In both cases plasma shielding is sustained for the remainder of the pulse, with near total shielding for Te. 
The normal velocity of the evaporated Te atoms (Fig. \ref{target comparison}b) peaks quickly near 600 m/s and then decays because the fully shielded target is cooling down for the remainder of the laser pulse. The velocity of the Se and Cu increases more slowly with both reaching the maximum velocity just after the onset of plasma shielding for Se and at the end of the pulse for Cu.

The distinct plasma shielding behavior of the three metal plumes stems from differences in their total plasma absorption. Fig. \ref{n0alphaT} details the plume density, absorption coefficients, and plasma temperature as a function of distance from the target at 15 ns, the midpoint time of the laser pulse. The number density of neutral atoms $n^0$ and electrons $n_e$ are shown in Fig. \ref{n0alphaT}a. For the chalcogens, their low boiling points cause the densities of evaporated neutral species at the surface of the targets to be as high as $\sim$10$^{28}$ m$^{-3}$. The density of Cu$^0$ is three orders of magnitude lower than the chalcogens at the surface. The spatial dependence of the plasma absorption coefficients is shown in Fig. \ref{n0alphaT}b. 

The effective model shows early expansion dominated by $\alpha_{PI}$ and $\alpha_{eff}$, with $\alpha_{IB}$ being of significantly less importance. Particularly high relative $\alpha_{eff}$ values emerge for the Cu case. The plasma temperatures are shown in Fig. \ref{n0alphaT}c. As expected, the temperature peaks in each case near the position where the plumes have the greatest $\alpha_{PI}$ and $n_e$. The Se plume has a peak temperature of 1.48 eV at 19 $\mu$m from the target, Te peaks at 1.04 eV at 41 $\mu$m, and Cu at 0.47 eV at 12 $\mu$m. The kinetic energy of singly-charged ions at the expansion front can be evaluated using the position of the sharp temperature increase at the plasma front and the elapsed time of 15 ns. Te has the greatest energy at 26.6 eV followed by Se and Cu at 9.68 eV and 3.02 eV, respectively.
\begin{figure}[t!]
    \centering
    \includegraphics[width = 0.95\linewidth]{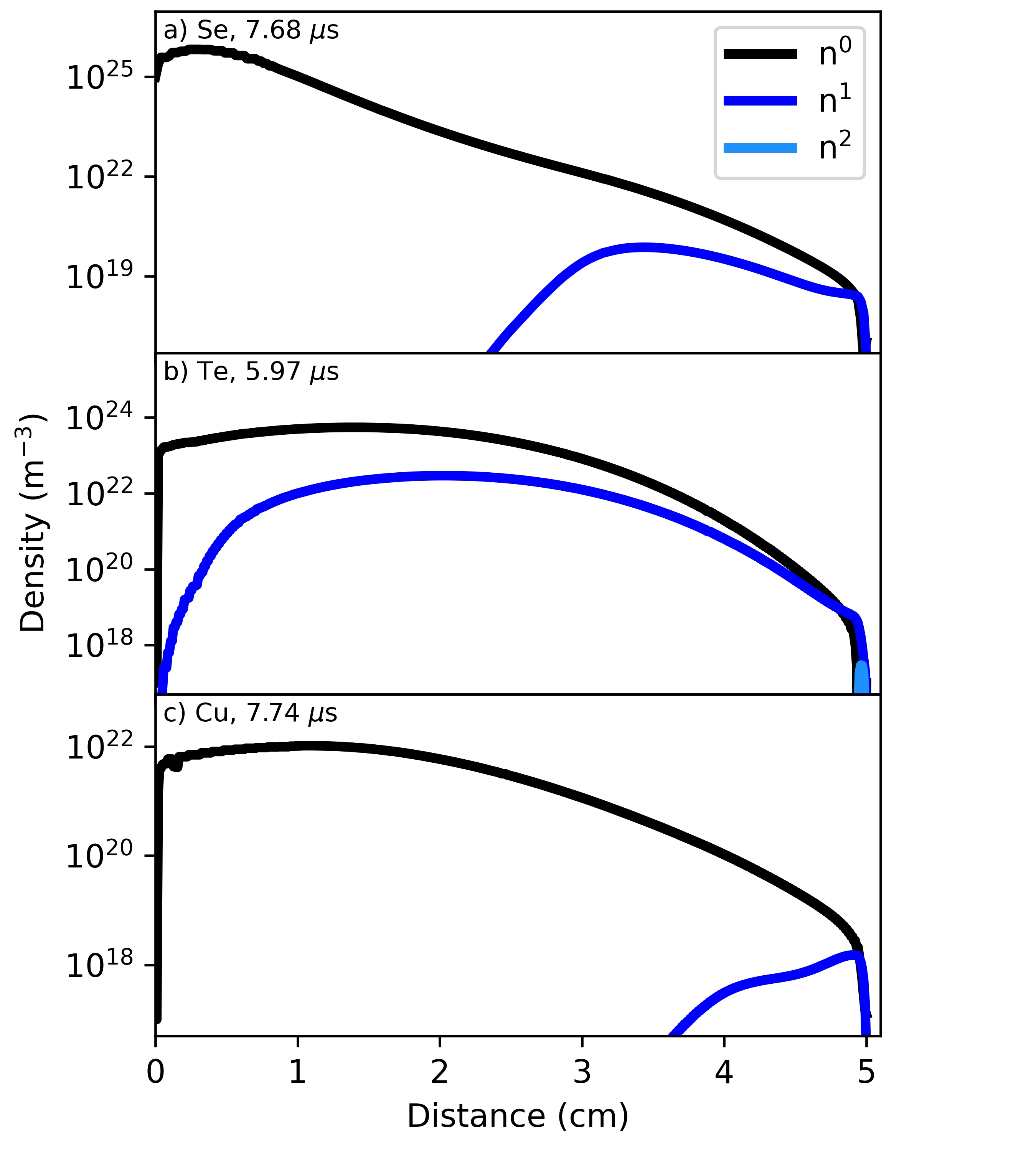}
    \caption{Qualitative 1D view for the long-term expansion for a) Se, b) Te, and c) Cu plumes. The Se and Te plasmas show greater ionization of their interiors while Cu ions are noted only in the frontal region of the plume. Ions at the expansion front have the greatest kinetic energy for the Te plume, followed by the less energetic expansions of Se and Cu.}
    \label{1d_expansion}
\end{figure}

A qualitative view of the long-term expansion of the plumes is obtained if simulations based on the 1D description are run for sufficiently long times as shown in Fig. \ref{1d_expansion}. We note that each plume reaches 5 cm at different times, 7.68 $\mu$s, 5.97 $\mu$s, and 7.74 $\mu$s for Se, Te, and Cu, respectively. The number densities are overestimated, since the plasma cannot expand in the transverse dimensions. Overall, the chalcogen plumes appear denser than the Cu plume, consistent with the early expansion profiles of Figs. \ref{target comparison} and \ref{n0alphaT}. The Se and Te expansions (Fig. \ref{1d_expansion}a,b) show increasing trend for ionization of their interiors. This is in contrast to the Cu plume, which has meaningful concentrations of ions only in the frontal region. For all three cases, the leading edge of the plume is fully ionized, with the Te plasma being somewhat unique due to the presence of a high Te$^{2+}$ number density at its leading edge for the used irradiation conditions.

\begin{figure*}[t]
    \centering
    \includegraphics[width = \linewidth]{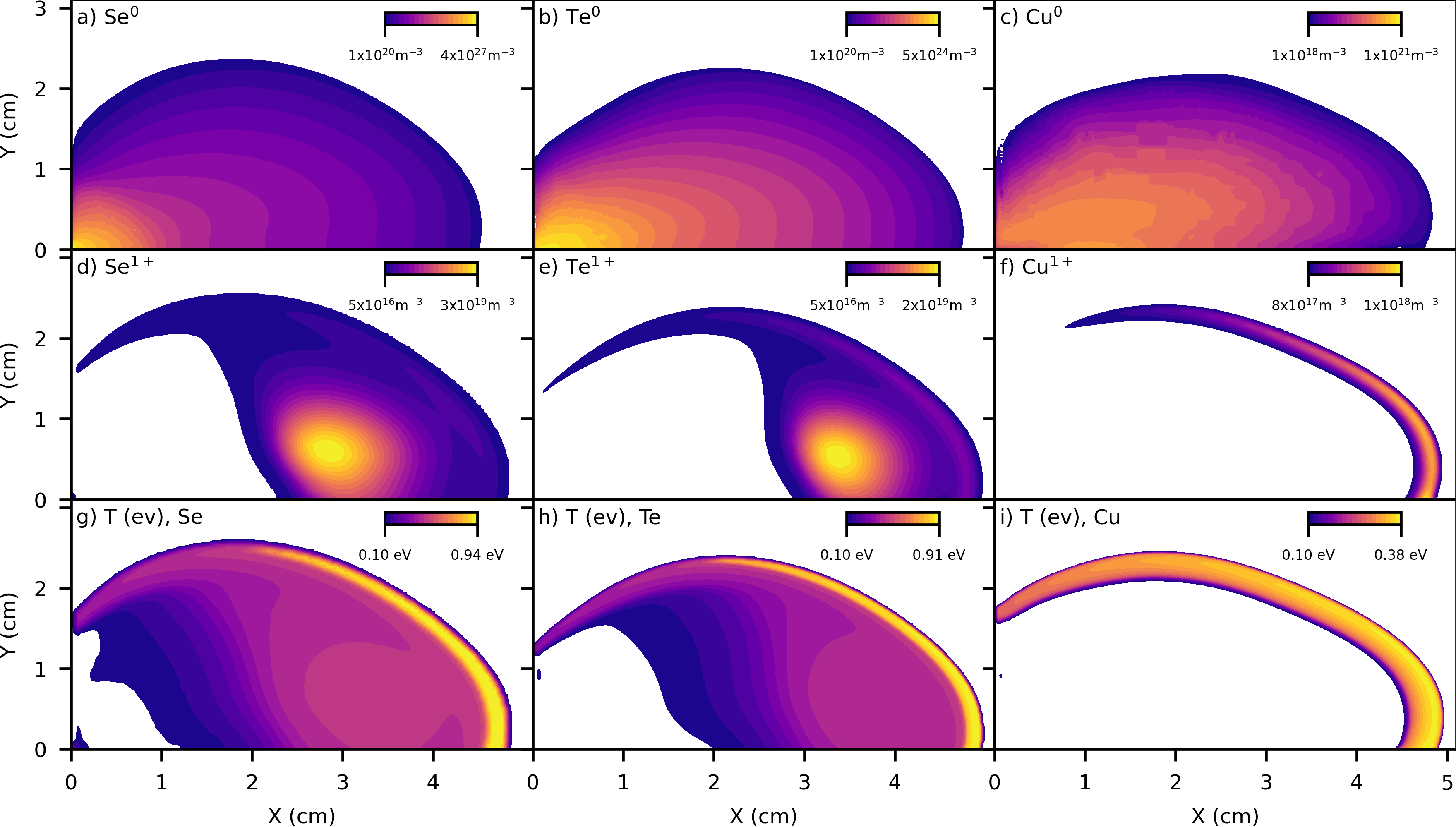}
    \caption{Contour plots for number density of neutrals (a,b,c), singly-charged ions (d,e,f), and plasma temperature (g,h,i) of experimentally-constrained 2D-axisymmetric simulations of Se, Te, and Cu plumes produced by ablation with a single KrF laser pulse of 4.0 J/cm$^2$ fluence and 1.5-mm circular spot diameter. Shown plume snapshots correspond to $t$ = 4.0 $\mu$s for Se and Te, and $t$ = 6.5 $\mu$s for Cu\textemdash when all expansion fronts have reached nearly 5 cm. Neutrals for Se, Te, and Cu (a,b,c) are plotted on a logarithmic color-scale due to spread of values over several orders of magnitude. d) Se$^{+}$ and e) Te$^{+}$ ions show well-defined peaks near a central bulge of the plasma expansion, while f) Cu$^{+}$ ions are only present at the leading edge of the plume and in densities lower by a factor of 10. The temperature of the chalcogen plumes (g,h), is $\sim$0.5 eV in the region of greatest ion density and peaks near 1 eV at the front. i) The temperature of the Cu plume is appreciable only at the expansion front with a maximum of 0.38 eV. Profiles shown are for values of $C_{eff}$ listed on Table \ref{tab:C_eff}, obtained as discussed in Section \ref{Sub:2}.}
    \label{2D_expansion}
\end{figure*}

A richer and expressly quantitative perspective on the long-distance expansion is furnished by the 2D-axisymmetric results shown in Fig. \ref{2D_expansion}. The distributions of neutral species for the chalcogens (Fig. \ref{2D_expansion}a,b) are heavily weighted towards the near-target region and fall off rapidly in all directions. Cu$^0$, on the other hand, is more homogeneous throughout the plume (Fig. \ref{2D_expansion}c). The number density of singly charged ions is also quite distinct for Se and Te, with well defined peaks roughly located near a central bulge in the plume (Fig. \ref{2D_expansion}d,e). This is in contrast to the Cu plume, which is dominated by Cu$^{0}$, with Cu$^+$ ions found exclusively at the leading edge of the expansion (Fig. \ref{2D_expansion}f). 

Consistent with the greater degree of ionization of the interior of their plumes, the temperature profiles of Fig. \ref{2D_expansion}g-i reveal that the chacogen plasmas are hotter throughout. Their temperatures are approximately 0.5 eV on the front half of the plume and near 1 eV at the leading edge, where there is an appreciable concentration of doubly ionized species. The concentrations of Se$^{2+}$ and Te$^{2+}$ (not shown) have peak densities of 5$\times$10$^{17}$ m$^{-3}$ and 1$\times$10$^{18}$ m$^{-3}$, respectively. Doubly ionized species only exist in $\sim$1 eV temperature regions and are therefore absent in the Cu plume, which is cooler than the chalcogens, reaching its highest value of 0.38 eV only at the expansion front. 

Fig. \ref{2D_expansion} illustrates the key advantage of multidimensional simulations: plasma processes are allowed to couple to the translational degrees of freedom that are perpendicular to the direction normal to the target. The most obvious benefit is bringing the predicted number densities into much closer agreement with experiments. We note that although the 1D density vs. distance profiles (Fig. \ref{1d_expansion}) show good qualitative correspondence with the 2D-axysimmetric plumes along the axis of symmetry (Fig. \ref{2D_expansion}a-f), the 1D number densities are not realistic. The maximum Te$^{+}$ density, for example, is 3$\times$10$^{22}$ m$^{-3}$ (1D) vs 2$\times$10$^{19}$ m$^{-3}$ (2D-axi). The lower densities brought about by plume expansion in the directions transverse to the symmetry axis are in significantly closer agreement with experimental density values, as can be inferred from data in Fig. \ref{TOF}, for instance.

The gains of the effective 2D-axisymmetric description allow quantitative studies of changes in the geometry of the expansion, which may vary from the quasi-spherical character of Knudsen-layer-only processes (Mach number $\sim 1$) to the highly forward-peaked profiles of hypersonic expansions \cite{KellyandMiotello}. The spatiotemporal dependence of the particle flux, the angular distribution of plasma species, and the effect of the laser spot size, can all be easily derived from these results to help guide experiments in materials growth and processing.

\section{Conclusion}
We have carried out 2D-axisymmetric simulations of laser-generated plasmas out to expansion distances pertinent to thin film deposition and modification. The simulations implement a laser-induced surface evaporation/plasma expansion model, which introduces an effective plasma absorption coefficient, as a stand-in for more complex ablation processes, whose parameters are generally unknown or difficult to compute. 
The simulation is constrained by experiments, by first determining the value of the effective absorption coefficient pre-factor ($C_{eff}$), that  causes the model to yield the essential features of density vs. ion TOF kinetic energy curves for a given material and irradiation conditions. Imposing this experimental constraint requires a broad parameter sweep, which is made practical by our fast computational framework.

We applied the effective simulation to the ablation of the chalcogens Se and Te by a laser pulse of 248 nm wavelength and $\sim$25 ns pulse duration. Under the same irradiation conditions (4.0 J/cm$^2$ fluence and 1.8 mm$^2$ circular spot area), the long-distance chalcogen plasmas are predicted to differ substantially from commonly studied Cu plumes. Their spatial gradients of plasma density are up to three orders of magnitude steeper than for Cu. Furthermore, their ion spatial distributions have central bulges that are quite distinct from the edge-only ionization of Cu plasmas. This correlates with the hotter temperature profiles of Se and Te ($0.50-1.0$ eV) in relation to those of Cu ($< 0.38$ eV). Prediction of these long-distance plasma properties, along with other characteristics readily obtainable from the simulation, such as angular distribution of species, spot size dependencies, and spatiotemporal variations of plasma parameters, can guide the use of chalcogen laser plasmas into new materials synthesis and modification regimes. Moreover, since the simulations are extendable to plasmas comprising additional multiple chemical species, they may also have future impact in the creation of binary, ternary, and quaternary compounds for which laser synthesis has a definite exploratory advantage over other methods.

\section*{Acknowledgments}
This work was supported in part by the National Science Foundation (NSF) EPSCoR RII-Track-1 Cooperative Agreement OIA-1655280 and by a grant of high performance computing resources and technical support from the Alabama Supercomputer Authority (ASA). SBH acknowledges graduate fellowship support from the National Aeronautics and Space Administration (NASA) Alabama Space Grant Consortium (ASGC) under award NNX15AJ18H. Any opinions, findings, and conclusions or recommendations expressed in this material are those of the authors and do not necessarily reflect the views of NSF, ASA, or NASA.

\bibliographystyle{apsrev}
 \typeout{}
\bibliography{main}

\end{document}